\documentclass[superscriptaddress,prd,onecolumn,aps,amssymb,nofootinbib]{revtex4}

\usepackage{latexsym}
\usepackage{graphicx}
\usepackage{amsmath}
\usepackage{amsfonts}
\usepackage{lscape}
\usepackage{dcolumn}
\usepackage{bm}

\begin{document}

\title{The Construction of Sudden Cosmological Singularities}
\author{John D. Barrow$^{1},$ S. Cotsakis$^{2}$ and A. Tsokaros$^{3}$ \\
$^{1}$DAMTP, Centre for Mathematical Sciences, \\
Cambridge University, Cambridge CB3 0WA, UK \\
$^{2,3}$ DICSE, University of the Aegean, Karlovassi 83200, Samos,
Greece}

\begin{abstract}
\noindent Solutions of the Friedmann-Lema\^{\i}tre cosmological equations of general
relativity have been found with finite-time singularities that are
everywhere regular, have regular Hubble expansion rate, and obey the
strong-energy conditions but possess pressure and acceleration singularities
at finite time that are not associated with geodesic incompleteness. We show
how these solutions with sudden singularities can be constructed using
fractional series methods and find the limiting form of the equation of
state on approach to the singularity.

\end{abstract}

\maketitle

\noindent In standard Friedmann-Lema\^{\i}tre cosmological models a fluid is placed in
a homogeneous and isotropic spatial geometry whose dynamics is then
determined by two independent Einstein equations for three unknown
time-dependent functions, the Friedmann metric
scale factor, $a(t)$, the fluid density, $\rho (t)$, and fluid pressure, $%
P(t)$, respectively (with units chosen with $c=8\pi G=1$):
\begin{eqnarray}
\frac{\dot{a}^{2}+k}{a^{2}} &=&\frac{\ \rho }{3}  \label{eq:FRWrho} \\
-2\frac{\ddot{a}}{a}-\frac{\dot{a}^{2}+k}{a^{2}} &=&\ P\ .  \label{eq:FRWp}
\end{eqnarray}%
If an equation of state $P=f(\rho )$ is chosen, the system closes and the
two remaining unknown functions are determined. Physically reasonable
equations of state, like those for perfect fluids, produce well behaved
expansion factors with behaviours that offer simple Newtonian
interpretations. However, the challenge of providing a compelling
explanation for the observed acceleration of the universe in terms of a
'dark energy' fluid has led to an exploration of other less familiar
equations of state \cite{jdb3}, motivated by the form of bulk viscous
stresses \cite{visc}, situations were $dP/d\rho $ is not everywhere
continuous, or scenarios in which there is no equation of state at all.
These introduce quite different possibilities into the solution space of
Friedmann universes and can produce unexpected types of finite-time
singularity which are far softer than the curvature singularities studied
previously \cite{tipl} in connection with geodesic incompleteness \cite%
{laz,HE}.

Barrow \cite{jdb04a,jdb04b} identified a whole new class of
pressure-driven singularities that keep the scale factor, $a$, expansion
rate, $\dot{a}/a$, and the density, $\rho $, finite while the pressure, $P$,
blows up at a finite time despite the energy conditions $\rho >0$ and $\rho
+3P>0$ holding. Their status as stable solutions of the classical Einstein
equations in the presence of small scalar, vector and tensor perturbations
has been studied in a gauge covariant formalism \cite{lip}, and they have
also been found to be stable against quantum particle production processes
\cite{fab}. They have been investigated in a number of different
cosmological scenarios and their behaviour has been classified in the light
of other types of finite-time singularity that can arise in isotropic and
anisotropic cosmologies \cite{bt,rev,jdb4,odin,dab}. The formal definitions
of these singularities has been discussed by Lake \cite{lake} in the case
where infinities occur in second derivatives of $a(t)$, but similar examples
exist where the sudden singularity occurs in higher derivatives \ of $a(t)$
and no energy conditions are threatened \ \cite{jdb04b,bt}. Here, we shall
provide an alternative systematic way of constructing them.

A sudden singularity will be said to arise everywhere at comoving proper
time $t_{s}$ in a Friedmann universe expanding with scale factor $a(t)$ if
\[
\lim_{t\rightarrow t_{s}}a(t)=a_{s}\neq 0,\qquad
\lim_{t\rightarrow t_{s}}\dot{a}(t)=\dot{a}_{s}<\infty ,\qquad
\lim_{t\rightarrow t_{s}}\ddot{a}(t)=\infty,
\]%
for some $t_{s}>0$. From the last requirement we have that, for all constant
$M>0$, there exists a constant $\epsilon >0$ such that for $|t-t_{s}|<\epsilon$
and $t<t_{s}$, we have
\begin{equation}
\ddot{a}(t)>M. \label{eq:ddota}
\end{equation}
By integrating the differential inequality (\ref{eq:ddota}), we have
\[
0<\int_{t}^{t_{s}}Mdt<\int_{t}^{t_{s}}\ddot{a}dt=\dot{a}_{s}-\dot{a}(t),
\]%
and with a further integration we see that for every positive number $M$ we
can find a sufficiently small left neighbourhood of $t_{s}$ such that for
every $t$ in that neighborhood the scale factor satisfies
\[
a(t)>a_{s}-\dot{a}_{s}(t_{s}-t)+\frac{M}{2}(t_{s}-t)^{2}.
\]%
For this condition to be satisfied in the neighborhood of $t_{s}$, we must
have
\begin{equation}
a(t)=a_{s}-\dot{a}_{s}(t_{s}-t)+C(t_{s}-t)^{n}+\cdots  \label{eq:BS}
\end{equation}%
where $C$ is a constant that depends on $M$, and \textit{necessarily} we $\ $%
have $1<n<2$.

If $n$ is a rational, the expansion factor can be written as a
Puiseux series of the form
\[
a(t)=\sum_{i=0}^{\infty }a_{i}(t_{s}-t)^{i/s}
\]%
where $s>0$ is a natural number.

We can study these kinds of singularities
by the dominant balance method \cite{scot}, wherein a dynamical system $%
\mathbf{\dot{x}}=\mathbf{f(x)}$ is split into dominant and subdominant terms
$\mathbf{f(x)}=\mathbf{f^{(0)}(x)}+\mathbf{f^{(sub)}(x)}$ and then we search
for solutions of the dominant dynamical system $\mathbf{\dot{x}}=\mathbf{%
f^{(0)}(x)}$ in the form of a Puiseux or $\Psi $ series. In our case,
setting $x=a,\ y=\dot{a}$ we get from (\ref{eq:FRWp})
\begin{equation}
\dot{x}=y,\qquad \qquad \dot{y}=-\frac{k+y^{2}+x^{2}P}{2x},  \label{eq:DynSys}
\end{equation}
and we require a dominant solution $x=a_{s}$ and $y=\dot{a}_{s}$ both
finite, with pressure being infinite. Clearly, these two conditions cannot
be simultaneously satisfied.

The root of the problem is that the dominant
balance method looks for a singularity of the variable
\[
\mathbf{x}=(\alpha _{1}(t_{s}-t)^{p_{1}},\alpha _{2}(t_{s}-t)^{p_{2}},\cdots),
\]%
(i.e. at least one of the $p_{i}<0$) but we want $\mathbf{x}$ to be finite
and $\mathbf{\dot{x}}$ (and so $\ddot{a}$) to have a singularity since the
pressure term appears in the higher derivative ($\ddot{a}$) equation (\ref%
{eq:FRWp}). Despite this structure, the general form of the pressure which
leads to this singularity can be found by looking for solutions of (\ref%
{eq:DynSys}) of the form
\begin{equation}
x(t)=\sum_{i=0}^{\infty }c_{1i}\ (t_{s}-t)^{i/s},\qquad
y(t)=\sum_{i=0}^{\infty }c_{2i}\ (t_{s}-t)^{i/s},  \label{eq:DynSysSol}
\end{equation}%
with $s>1$. We want
\[
c_{10}=\alpha \neq 0, \qquad c_{20}=\beta \neq 0,
\]%
to hold. In this case, the pressure takes the form
\begin{equation}
P=(t_{s}-t)^{h}\sum_{i=0}^{\infty }p_{i}\ (t_{s}-t)^{i/s},  \label{eq:PresSer}
\end{equation}%
and for $P$ to blow up at $t_{s}$, we must have $h<0$. The recursion
relations for the $c_{ij}\ $are
\[
c_{1s}=0, \qquad 0<i<s,\qquad c_{2i}=-\frac{s+i}{s}c_{1(s+i)},
\]%
and
\begin{eqnarray*}
&&-\frac{2(s+1)\alpha }{s^{2}}c_{1(s+1)}(t_{s}-t)^{-(s-1)/s}-\frac{%
4(s+2)\alpha }{s^{2}}c_{1(s+2)}(t_{s}-t)^{-(s-2)/s}-\cdots \\
&&-\frac{2(s-1)(2s-1)\alpha }{s^{2}}c_{1(2s-1)}(t_{s}-t)^{-1/s}-(4\alpha
c_{1(2s)}+\beta ^{2}+k)+ \\
&&+\frac{2(s+1)}{s^{2}}[\beta (s+1)c_{1(s+1)}-\alpha
(2s+1)c_{1(2s+1)}](t_{s}-t)^{1/s}+ \\
&&+\left[ \frac{2(s+2)^{2}\beta }{s^{2}}c_{1(s+2)}-\frac{(s+1)(s+3)}{s^{2}}%
c_{1(s+1)}^{2}-\frac{4(s+1)(s+2)\alpha }{s^{2}}c_{1(2s+2)}\right]
(t_{s}-t)^{2/s} \\
&&+\cdots =\alpha^2 p_{0}(t_{s}-t)^{h}+\alpha
^{2}p_{1}(t_{s}-t)^{h+1/s}+\alpha ^{2}p_{2}(t_{s}-t)^{h+2/s}+\cdots
\end{eqnarray*}%
By balancing the terms in this equation we can determine the series
expansion of the pressure that will lead to a specific sudden singularity.
For example, if $n=3/2$ in \cite{jdb04a}
\[
a(t)=\left( \frac{t}{t_{s}}\right) ^{q}(a_{s}-1)+1-\left( 1-\frac{t}{t_{s}}%
\right) ^{n},
\]%
we get for the pressure:
\[
P=\frac{3}{2\alpha t_{s}^{3/2}}(t_{s}-t)^{-1/2}-\frac{4\alpha \alpha
_{2}+\beta ^{2}+k}{\alpha ^{2}}-\frac{9\beta }{2\alpha ^{2}t_{s}^{3/2}}%
(t_{s}-t)^{1/2}+\cdots ,
\]%
with
\[
\alpha =\alpha_{s},\qquad a_{i}=(a_{s}-1)\binom{q}{i}\left( -\frac{1}{t_{s}}\right)^{i},
        \qquad \beta =-a_{1},
\]%
and the density has the form
\[
\rho =\rho _{s}+\frac{9\beta }{\alpha ^{2}t_{s}^{3/2}}(t_{s}-t)^{1/2}+\cdots
,\qquad \qquad \rho _{s}=\frac{3(k+\beta ^{2})}{\alpha ^{2}}\ .
\]%
We note here that the asymptotic form of the equation of state has the
simple form
\[
\lim_{t\rightarrow t_{s}}(P(\rho -\rho _{s}))=\frac{27\beta }{2\alpha
^{3}t_{s}^{3}},
\]%
and so the fluid behaves like a Chaplygin gas on approach to the sudden
singularity. Hence, we see that these formal methods for the analysis of
finite-time singularities in dynamical systems allow us to construct
cosmological singularities of the sudden type.

In conclusion, we have shown how to construct evolutionary behaviours
meeting the conditions for a sudden cosmological singularity using analytic
techniques with fractional power series employed to study finite-time
singularities in differential equations.

\end{document}